\documentclass[final,3p,times,twocolumn]{elsarticle}
 \biboptions{comma,sort&compress}
\usepackage{here}
 \usepackage{graphicx}
  \usepackage{epsfig}

\def\nin{\noindent}
\def\beq{\begin{equation}}
\def\eeq{\end{equation}}
\def\bea{\begin{eqnarray}}
\def\eea{\end{eqnarray}}

\def\pythia{{\sc Pythia}}

\def\powheg{{\sc Powheg}}

\begin{document}

\begin{frontmatter}

\title{Higgs as a gluon trigger} 

\author[label1]{P. Cipriano}
  \address[label1]{Deutsches Elektronen Synchrotron,  D 22603 Hamburg }  

\author[label1]{S. Dooling}

\author[label1]{A. Grebenyuk}

\author[label1]{P. Gunnellini}

 \author[label2,label3,label4]{F. Hautmann}
  \address[label2]{University of Sussex, Physics $\&$ Astronomy, Brighton BN1 9QH}
  \address[label3]{University of Oxford,  Physics  Department, Oxford OX1 3NP}
 \address[label4]{Rutherford Appleton Laboratory, Chilton OX11 0QX}

 \author[label1,label5]{H. Jung}
\address[label5]{Universiteit Antwerpen, Elementaire Deeltjes Fysica, B 2020 Antwerpen}

 \author[label1]{P. Katsas}

\begin{abstract}
\noindent
In the forthcoming high-luminosity phase at the LHC many of the  
  most interesting  measurements    for precision QCD studies
 are hampered  by conditions of    large   pile-up, particularly  at 
 not very high transverse momenta.   We  study observables based on 
 measuring 
  ratios of color-singlet currents    via Higgs boson and Drell-Yan production, 
which may be accessed also  at  large pile-up,    and  used  for an 
   experimental program on QCD physics of 
     gluon fusion processes  in the  LHC  high-luminosity  runs.  
We   present  results of  Monte Carlo   calculations   for a few specific examples.  
\vskip -9.6cm
\hspace*{13.5 cm} {  DESY 13-139 } 
\vskip 9.2cm
\end{abstract}

\end{frontmatter}

\nin   
 The  observation 
of the Higgs boson by the ATLAS and CMS experiments~\cite{hgs-obs} 
marks the beginning of a revolutionary era in high-energy physics.  It affects   
 profoundly  the  paradigms by which we 
 define  the limits of our knowledge 
 on the nature of interactions of elementary particles.   
 This   observation   gives us confidence in the  
physical  picture  of  
 fundamental interactions  encoded by the Standard Model (SM)  Lagrangian 
and provides us with guidance  in the search for   its   generalizations. 

The electro-weak sector of the SM and 
the  nature of  electro-weak symmetry breaking will be explored in detail 
in the coming years of operation   of the LHC by measuring properties of 
the observed boson~\cite{eurstra}.  In this note we remark   that  the  observation of the 
Higgs   boson  opens  up 
 the possibility  of a rich  experimental program 
 in the strong-interaction sector of the SM as well.  In particular, we  propose  
that a program of QCD measurements at high luminosity 
can be  carried out 
at the LHC by using the Higgs  boson as a trigger,   focusing     on 
QCD gluonic processes at high mass scales.

Classic collider probes of QCD in $e^+ e^-$ annihilation, deep inelastic 
$ ep$  scattering (DIS), Drell-Yan production (DY) all involve color-singlet currents 
which couple to quarks. With the Higgs, for the first time, LHC experiments will 
probe QCD by a color-singlet current which, in the heavy top limit,  couples to gluons. 
The physics of  gluon fusion processes  can 
 be explored from a 
  new perspective  compared to experimental investigations over the past three 
decades.  As  illustrated below, 
 we propose  measuring  systematically differences  of differential 
distributions for Higgs  and    Drell-Yan final states. 
This comparison  allows  one to access experimentally distinctive QCD features of 
gluon fusion physics. 

In the next    high-intensity phase  at the LHC, one faces high pile-up conditions 
leading to large numbers of overlaid events.  In these conditions 
many of the    most interesting  measurements    for precision QCD studies, 
  particularly   for  not very high  transverse momenta, 
  become extremely difficult - see e.g.~\cite{eurstra,eur-qcd}.     
Here we argue  that by  studying differences  of Higgs and  
Drell-Yan for  masses around 125 GeV  the effects of pile-up largely drop out. 
This offers  the possibility of a program of  QCD measurements  
of  great physics interest   in the high-luminosity runs of the LHC. 

In  this paper     we illustrate  this by Monte Carlo simulation  for  
three  specific examples:    the  ratio of Higgs 
vs. Drell-Yan $p_\perp$ spectra; the structure of the associated 
 underlying event    and charged-particle  multiplicities;  the 
scattering angle in the center-of-mass reference frame. 
  These      
involve QCD physics both at high transverse momenta and at low 
transverse momenta, and    allow  one to study both 
high-$x$ and low-$x$ physical effects. 

We  contrast  the 
distinctive features of 
the Higgs trigger  with other  LHC 
 short-distance probes such as  
jets, heavy flavors,   vector boson pairs  which either couple  
perturbatively to color-octet and color-triplet sources 
on an equal footing,  or imply 
final-state color-charged particles,  or both.

We leave to    detailed 
phenomenological  investigations   the study of 
 the optimal channels 
to be used to  access 
gluon fusion and suppress Higgs production by 
 vector boson fusion and 
quark  annihilation;  of  the luminosity requirements for  
   reaching  sufficient statistics; of the different 
 treatment of pile-up  for  different channels.

Very recently the ATLAS collaboration has presented 
first  measurements of Higgs differential cross sections 
 based on the 2012 
dataset in the diphoton decay channel~\cite{atlas-conf-eps}.

\section*{Higgs vs. Drell-Yan}

\begin{figure}[hbt] 
\vspace{40mm}
\includegraphics{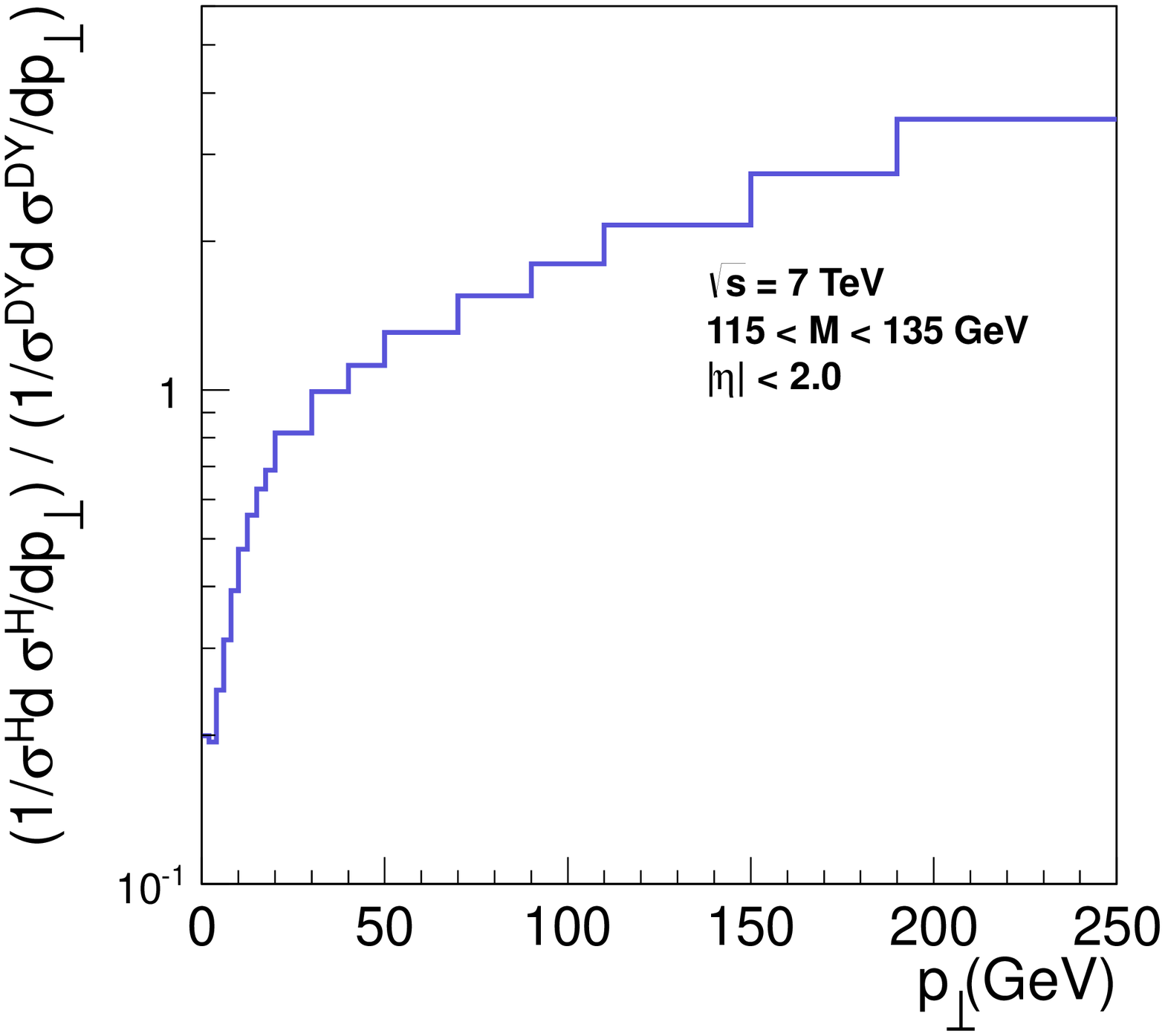}
\includegraphics{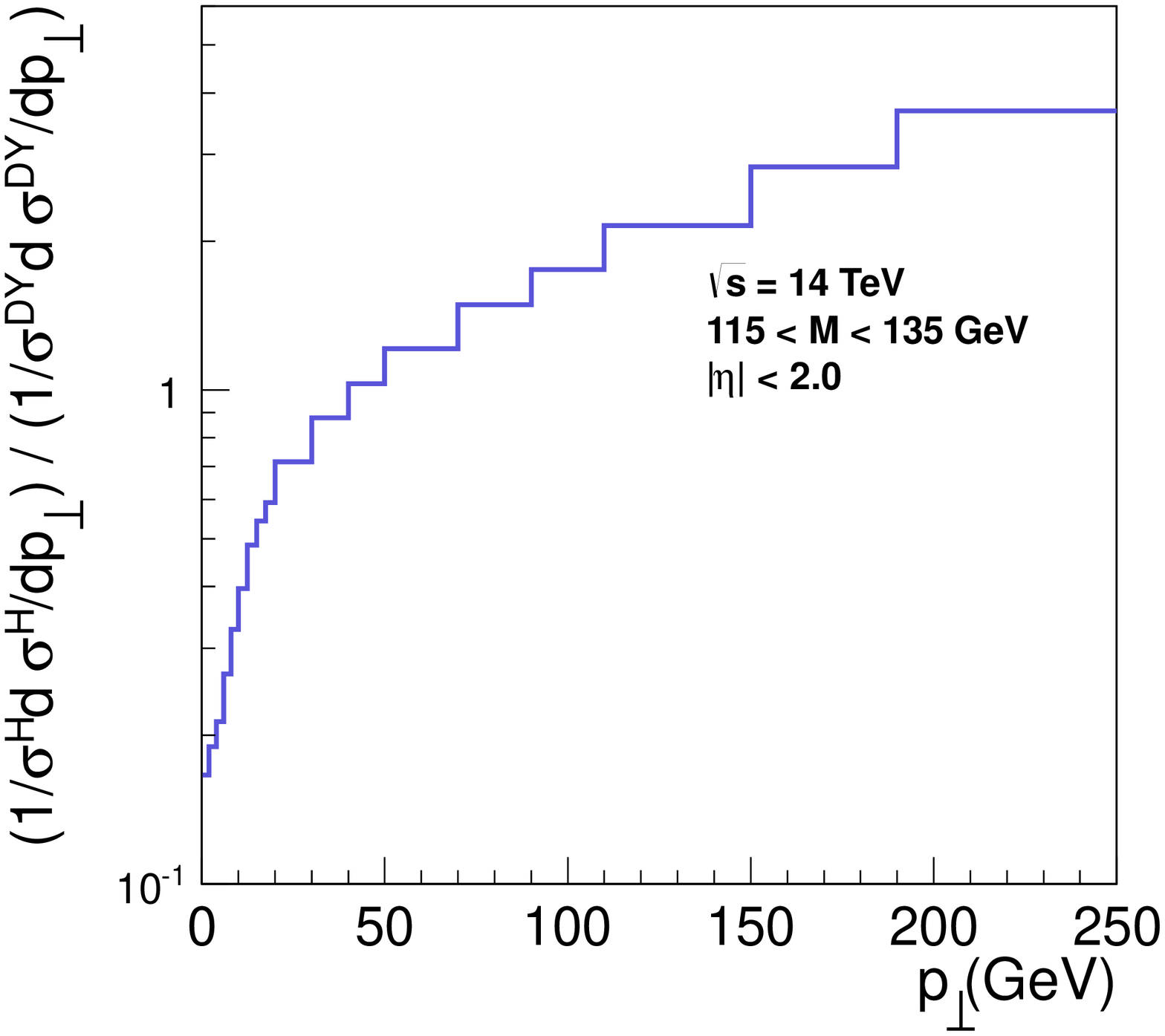}
\caption{\scriptsize Ratio of Higgs to DY spectra  versus $p_\perp$.}
\label{fig:fig-pt-ratio} 
\end{figure}

\nin Consider first 
  transverse momentum spectra  for Higgs bosons and for Drell-Yan (DY) 
pairs  in the invariant mass  range 115  GeV $ < M < $ 135 GeV.  
Transverse momentum spectra, comparing 
 Higgs  and $Z$ bosons,    
were  examined early on in~\cite{hin-novaes}. 
The transverse momentum spectra  
can be described  by  QCD factorization  in  the form 
\begin{equation} 
\label{ptsp-fact} 
d \sigma / d p_\perp = \int H \otimes S \otimes J_1 \otimes J_2  \;\; , 
\end{equation}   
decomposing the cross section  
 into hard ($H$), soft ($S$) and collinear-to-initial-states  ($J_1$, $J_2$) 
contributions -  see e.g.~\cite{jcc-fh-00} for  
analysis of how  this  decomposition     arises.  
\begin{figure}[hbt] 
\vspace{40mm}
\includegraphics{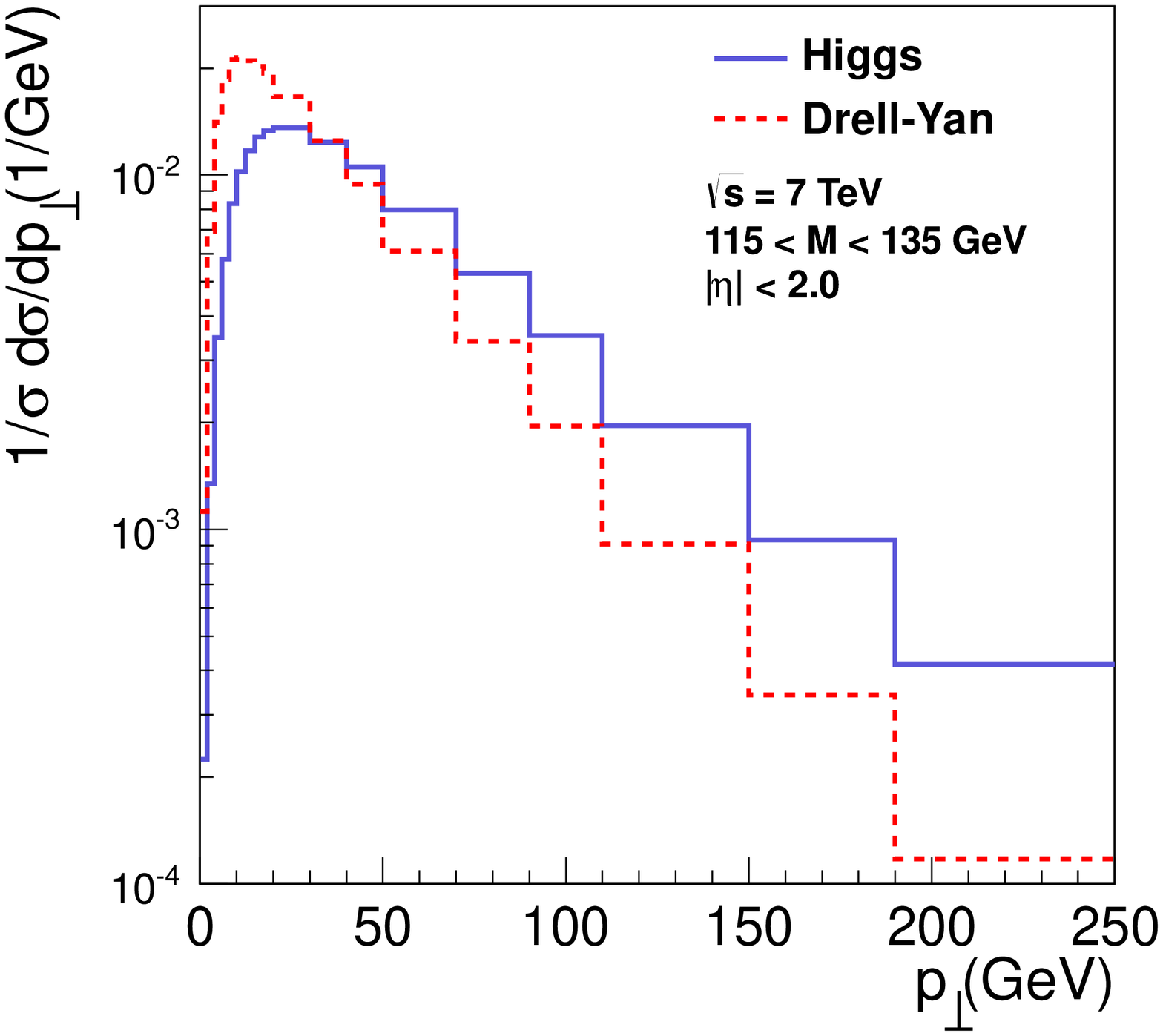}
\includegraphics{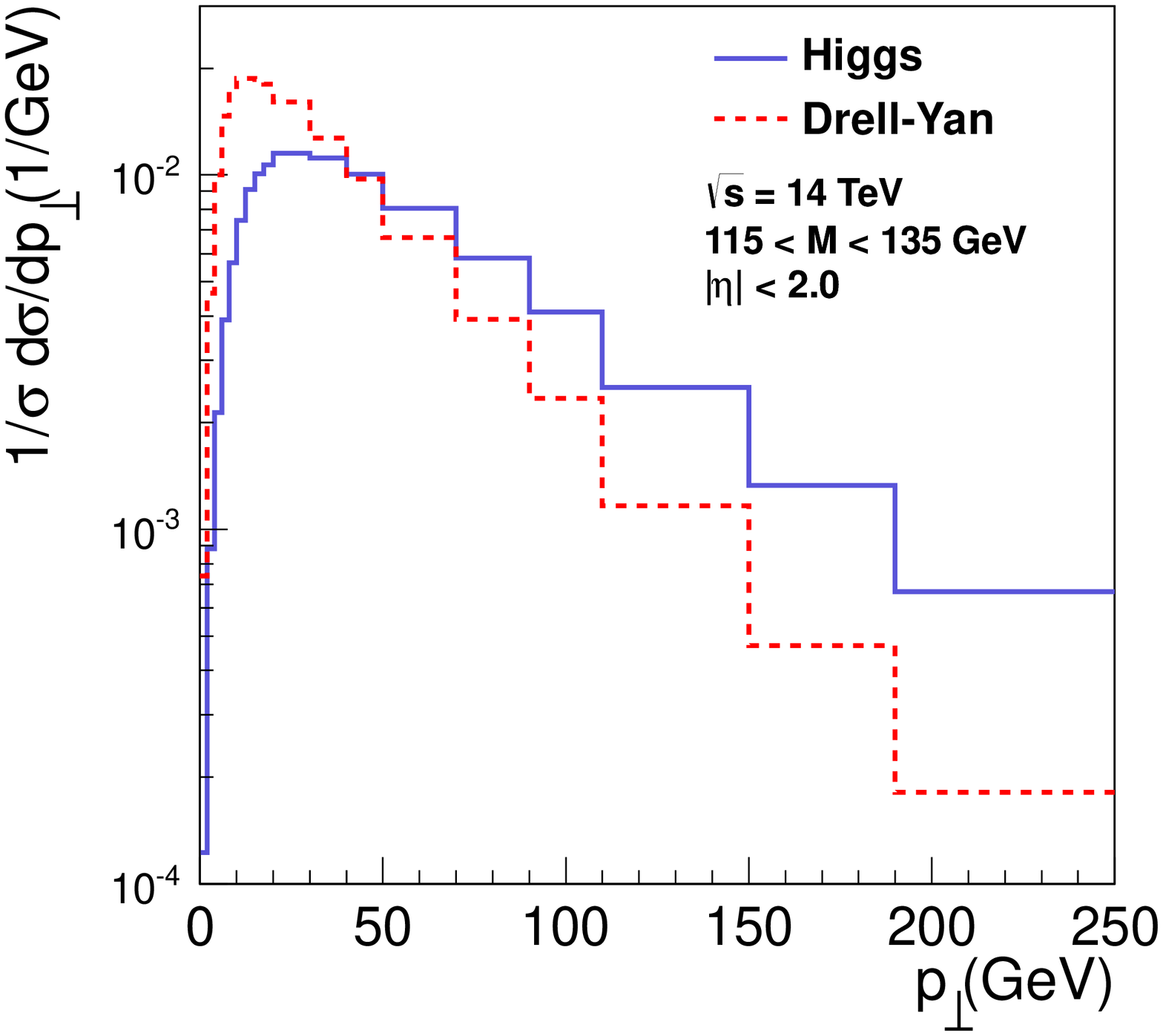}
\caption{\scriptsize Normalized transverse-momentum spectra 
for Higgs bosons and for Drell-Yan 
pairs.}
\label{fig:fig-pt} 
\end{figure} 
In Fig.~\ref{fig:fig-pt} we  show the result   of  Monte Carlo  
simulations for the  $ p_\perp $  spectra in the central region 
based on the next-to-leading-order (NLO) 
\powheg~\cite{powhegcit} event generator interfaced with 
\pythia~\cite{pythia6} shower, at $\sqrt{s} = 7 $ TeV and $\sqrt{s} = 14 $ TeV.   
In Fig.~\ref{fig:fig-pt-ratio} we plot the ratio of the Higgs and DY 
spectra at invariant mass 115  GeV $ < M < $ 135 GeV. 

The  $   p_\perp \ll     M  $  region of the spectrum  
measures 
 infrared  aspects of  the cross section in Eq.~(\ref{ptsp-fact}), i.e. 
i) the ratio of the gluon 
vs. quark Sudakov form factor (factor $S$ in Eq.~(\ref{ptsp-fact})), and ii) the 
evolution of the  collinear-to-initial-states  functions 
(factors $J_1$, $J_2$  in Eq.~(\ref{ptsp-fact})). 
In particular, gluon polarization terms 
$p_\perp^\mu p_\perp^\nu$ in gluon fusion, 
related to eikonal  polarizations at high energy~\cite{hgs02}, 
  give rise to distinctive 
radiation patterns from  
 initial-state functions in the Higgs case  -  see 
e.g.~\cite{mantry}.    
The $   p_\perp  \geq     M  $ region measures  the  ultraviolet function $H$ 
 in Eq.~(\ref{ptsp-fact})  and the features of   hard jets 
  recoiling against  Higgs  and DY pair. In particular,  the 
 leading-jet contribution to the  measured ratio 
  depends on the   $p_\perp$ distribution 
for  spin-1   vs. spin-1/2 exchange  and  on the  corresponding 
color emission probabilities.  Further  aspects on  jet recoil are discussed 
below in  the context of angular distributions.  

 In the large pile-up environment 
of the high-luminosity LHC runs, one has to deal with 
the  contribution of   large numbers of overlaid events. 
However, this contribution 
 cancels in the comparison   of Higgs to DY spectra at  fixed   invariant mass.   
Using this  comparison 
  one can  go to  low $p_\perp$ and  access 
  QCD effects in this region experimentally  
also at high pile-up.

Measurements  on gluon fusion   which  can be 
performed  using the Higgs  trigger  
open  a  new experimental  area. They 
may also be relevant  to interpret data for  other, more complex  processes, e.g. 
processes that depend on both quark and gluon channels on an equal footing, or involve 
color-charged particles in the final state. 

One such  example is 
given by top quark production.  This is often  studied as a 
process sensitive to gluonic initial states  at the LHC. 
 For   instance, 
the top quark $p_\perp$ spectrum~\cite{cms-top}  
receives contribution at low $p_\perp$
from the gluon Sudakov form factor and gluonic initial-state recoil 
analogous to those discussed above. 
However, since the final state   
 is not  a color-singlet current,  
the analysis of the $p_\perp$ spectrum is made more complex by  
 final   soft color  emission. 
 The Higgs case serves to 
single out  the   initial-state 
   contributions, including gluon 
 polarization effects. 

For observables more exclusive   than the cross section   in Eq.~(\ref{ptsp-fact}),     
e.g.  measuring the associated  jets,  full  QCD factorization formulas are still lacking.  
For   parton shower  event generators,    inclusive measurements are still useful 
 to  control  methods~\cite{plaetzer,prestel}  
  for  merging     parton showers and  matrix elements. 
Higgs vs. DY  studies 
similar to those considered above can be  done,   for instance,  
 in  boson + jet states, now  fixing, in addition to invariant mass, the jet  
transverse momentum or  rapidity.

\section*{Underlying events}

\nin    The structure of  
underlying events and color flows associated 
with Higgs boson final states 
was investigated long ago~\cite{doksjohgs}  
as a  possible method   to   
analyze    $gg \to H $  
 and $W W \to H$  production mechanisms. 
In the case of vector boson final states 
 it  was  pointed 
out~\cite{skands}  that the  treatment of parton showers, 
 and in particular of the recoils 
in the shower,  is essential for a proper description of $ W / Z $  spectra. 
This affects the amount of 
multi-parton interactions~\cite{zijl} needed to describe the 
events~\cite{skands,tunes}.  Analogous 
effects may be investigated for 
 gluonic showers~\cite{lonn-sjo,deak1006}   
in the case of  events associated with 
Higgs final states. 

\begin{figure}[hbt] 
\vspace{40mm}
\includegraphics{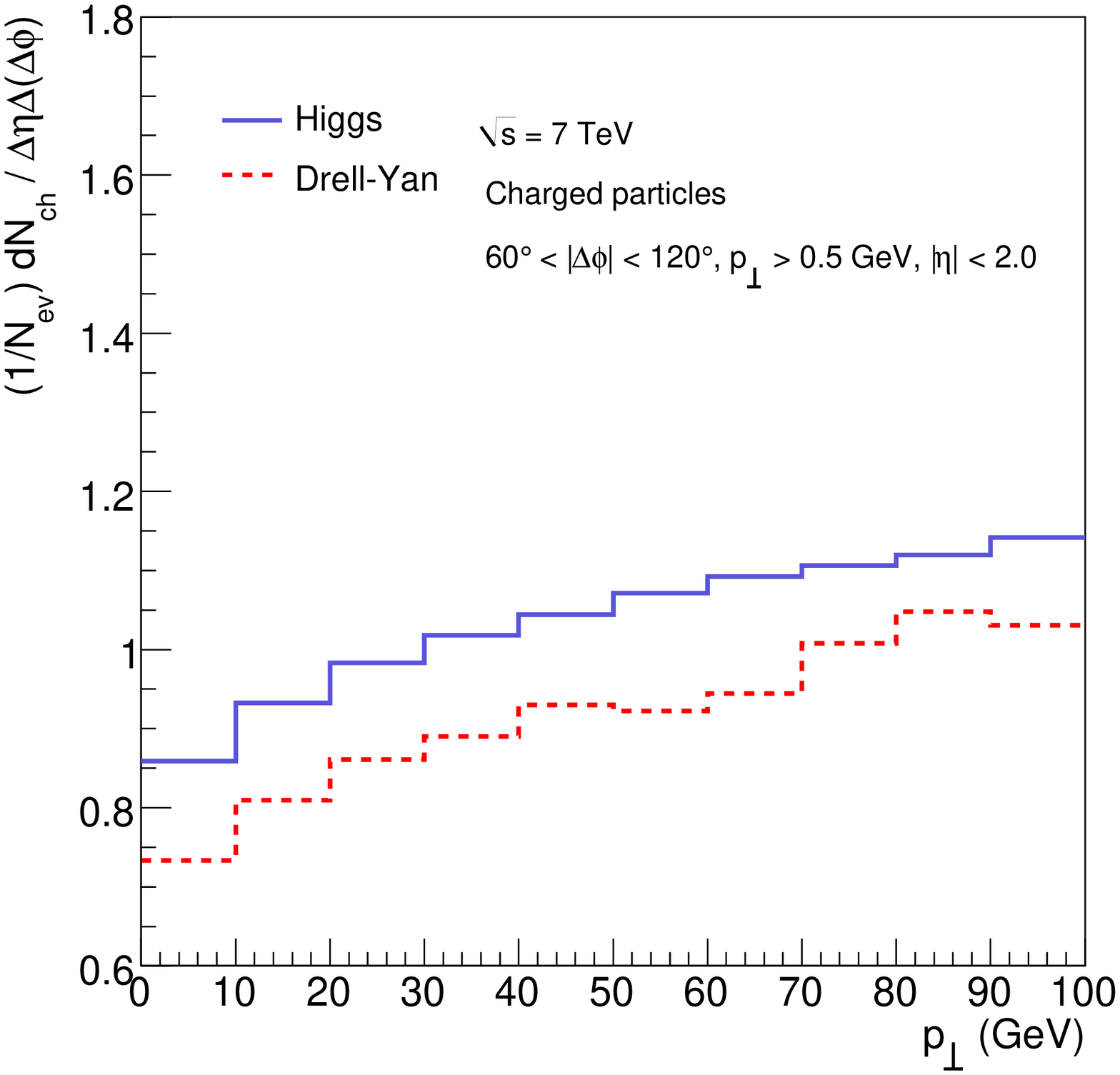}
\includegraphics{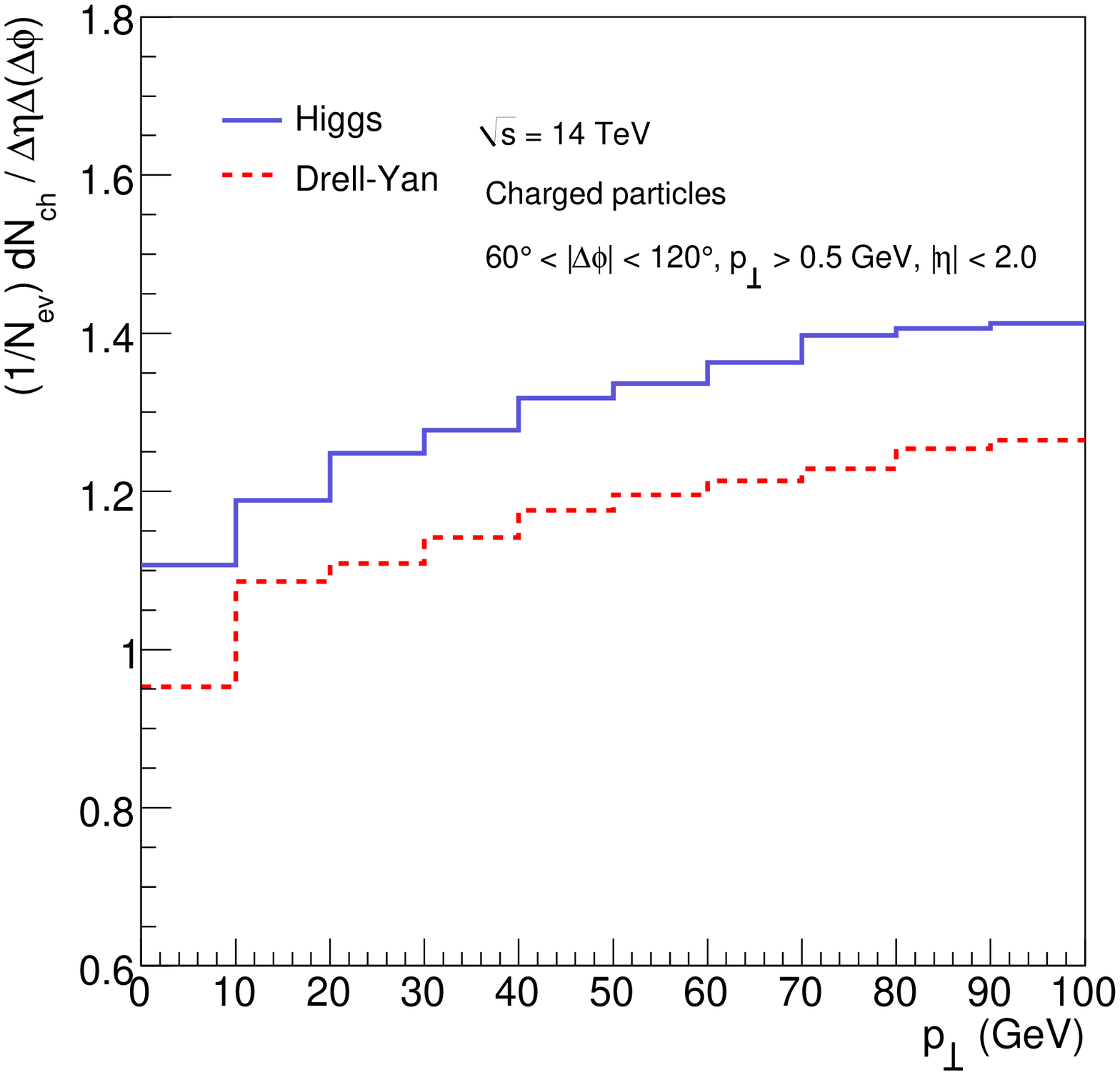}
\caption{\scriptsize Normalized 
charged-particle  average multiplicity in the transverse region  of the azimuthal plane 
versus  
Higgs transverse momentum  (solid blue line)  and DY transverse momentum (red dashed line).}
\label{fig:ue-mult1} 
\end{figure}

\begin{figure}[hbt] 
\vspace{40mm}
\includegraphics{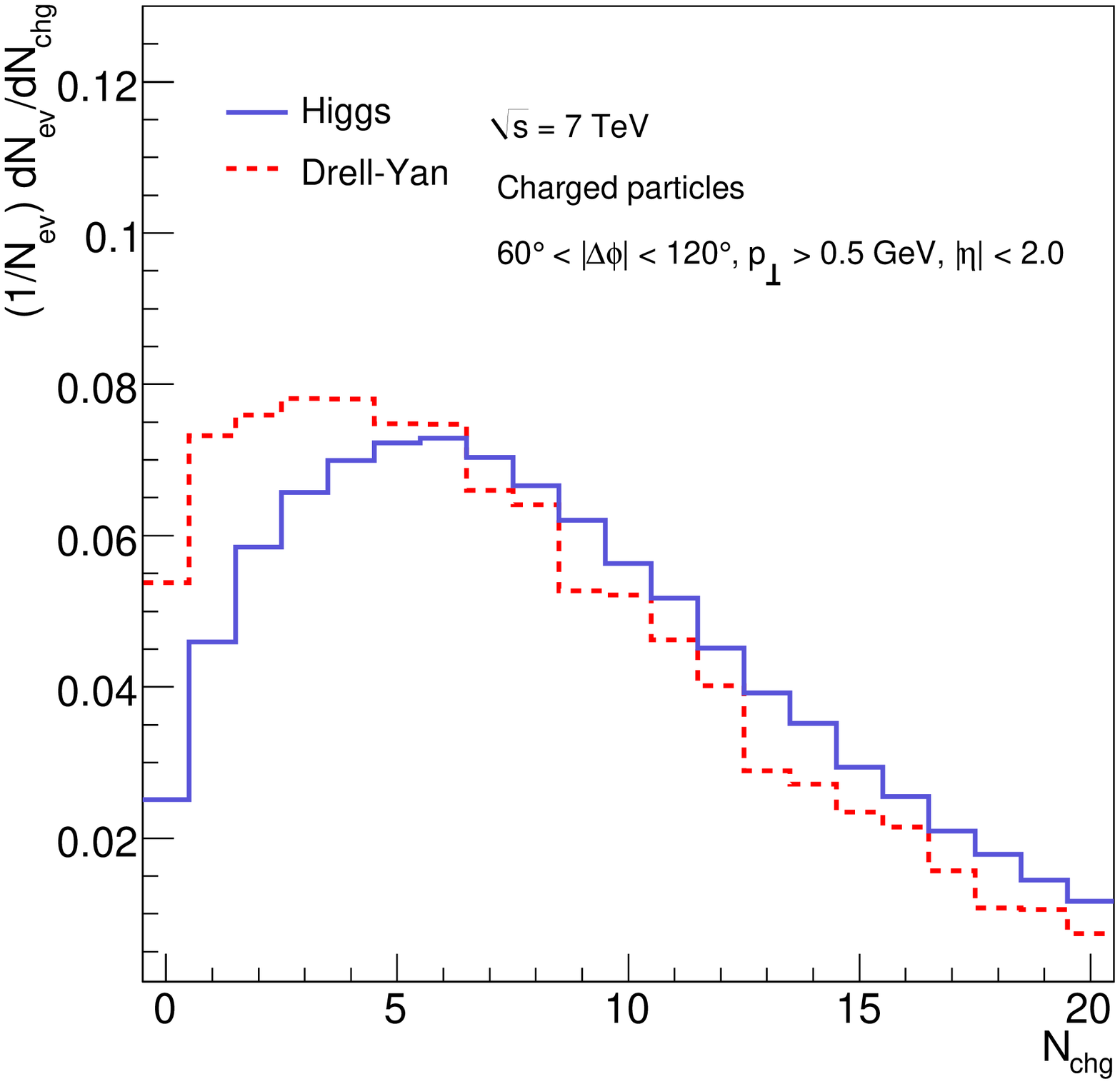}
\includegraphics{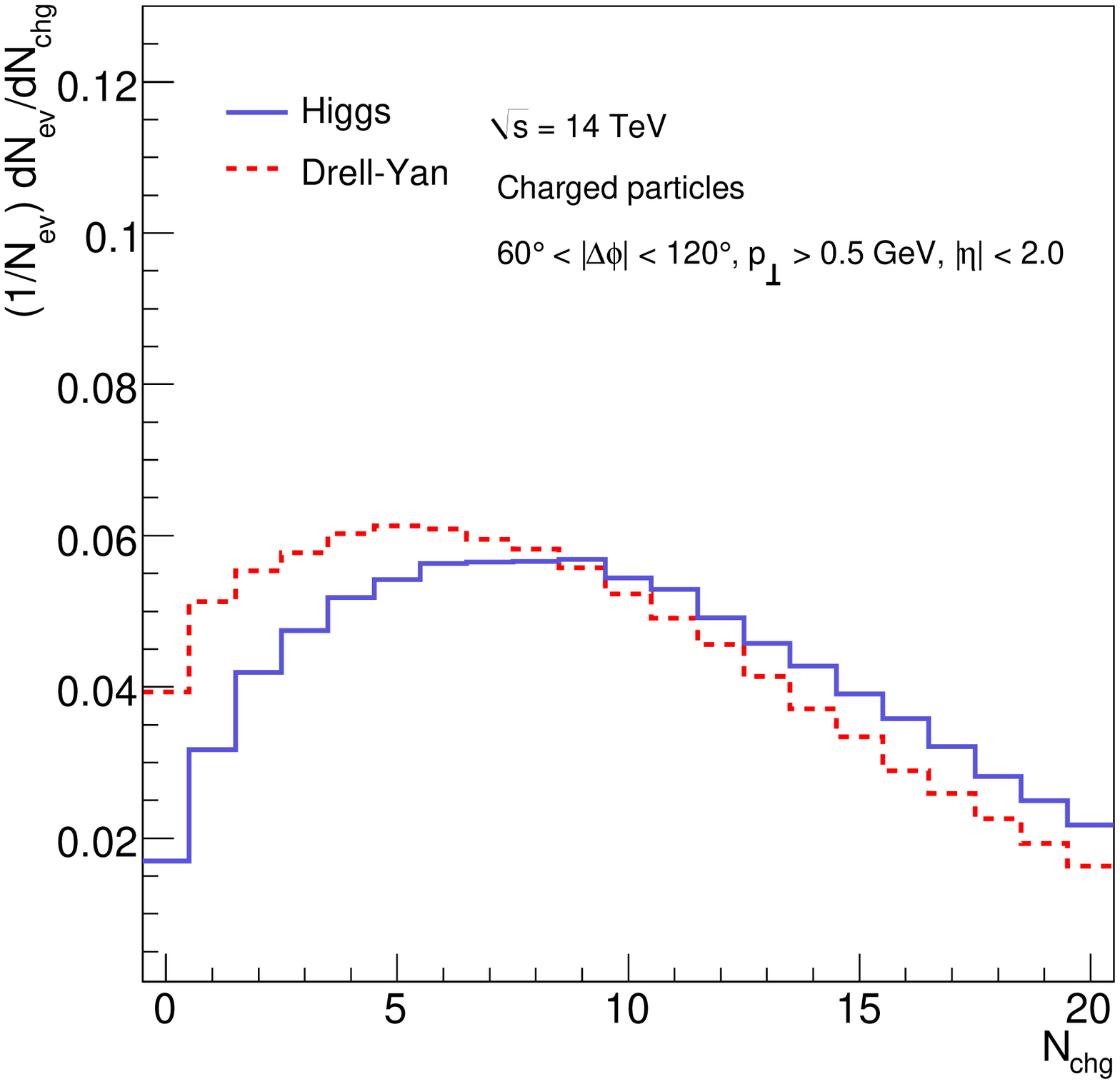}
\caption{\scriptsize 
Charged-particle   multiplicity  distribution 
in the transverse region  of the azimuthal plane in the 
Higgs  (solid blue line)  and Drell-Yan (red dashed line)  cases.}
\label{fig:ue-mult2} 
\end{figure}

We follow the    treatment~\cite{cms-ue} 
of underlying events in the azimuthal plane, 
with the direction of the Higgs momentum  and the   
DY pair momentum, respectively,   defining the origin in the azimuthal plane. 
In Figs.~\ref{fig:ue-mult1} and~\ref{fig:ue-mult2} 
  we show the result  of     NLO \powheg\ +  \pythia\    Monte Carlo calculations 
 for   charged-particle multiplicities associated with  Higgs  and DY. 
(Analogous calculations can be usefully performed for  multiplicities of mini-jets 
defined e.g.~as in~\cite{epjc12}.) 
We plot  the average multiplicity versus Higgs and DY $p_\perp$ (Fig.~\ref{fig:ue-mult1})
and the multiplicity distribution (Fig.~\ref{fig:ue-mult2}) in the 
transverse region of the azimuthal plane ($60^\circ < | \Delta \Phi | < 120^\circ$).

The distributions in the  Higgs case  are dominated by higher 
 multiplicities from gluon cascades.

Similarly to  the case of  the previous section,  
 the effects  of large number of overlaid events due to pile-up 
will be   reduced  if one  measures   
the difference between Higgs and DY underlying event distributions.

\section*{Angular distributions}   

\nin   
Besides  soft radiation from  underlying events,  
we 
 consider Higgs versus DY distributions 
 in the case of   hard  radiation 
accompanying the  heavy bosons,  for  example  boson + jet~\cite{eksbook}.  
For  Higgs production 
the angular distribution in the scattering angle $\theta^\ast$
of   the  boson-jet    center-of-mass  frame 
is characterized by  the scalar coupling to gluons  partially  canceling  
  the small-angle Coulomb singularity 
$ d \theta^{\ast 2} / \theta^{\ast 4}$   from 
 gluon scattering  - see e.g~\cite{hgs02}. 
The  Drell-Yan $\theta^\ast$   distribution is determined by  spin-1/2 exchange.   
Owing to the cancellation 
from the scalar coupling to gluons,   
 the  angular  distributions have  the same 
small-angle asymptotics in the Higgs and DY cases,  
despite the two processes occurring via spin-1 and spin-1/2 
exchange. 
 The $\theta^\ast \to 0 $ behavior  
thus tests  the   Higgs  spin  at the level of the 
 production cross section.

\begin{figure}[htbp]
\vspace{40mm}
\includegraphics{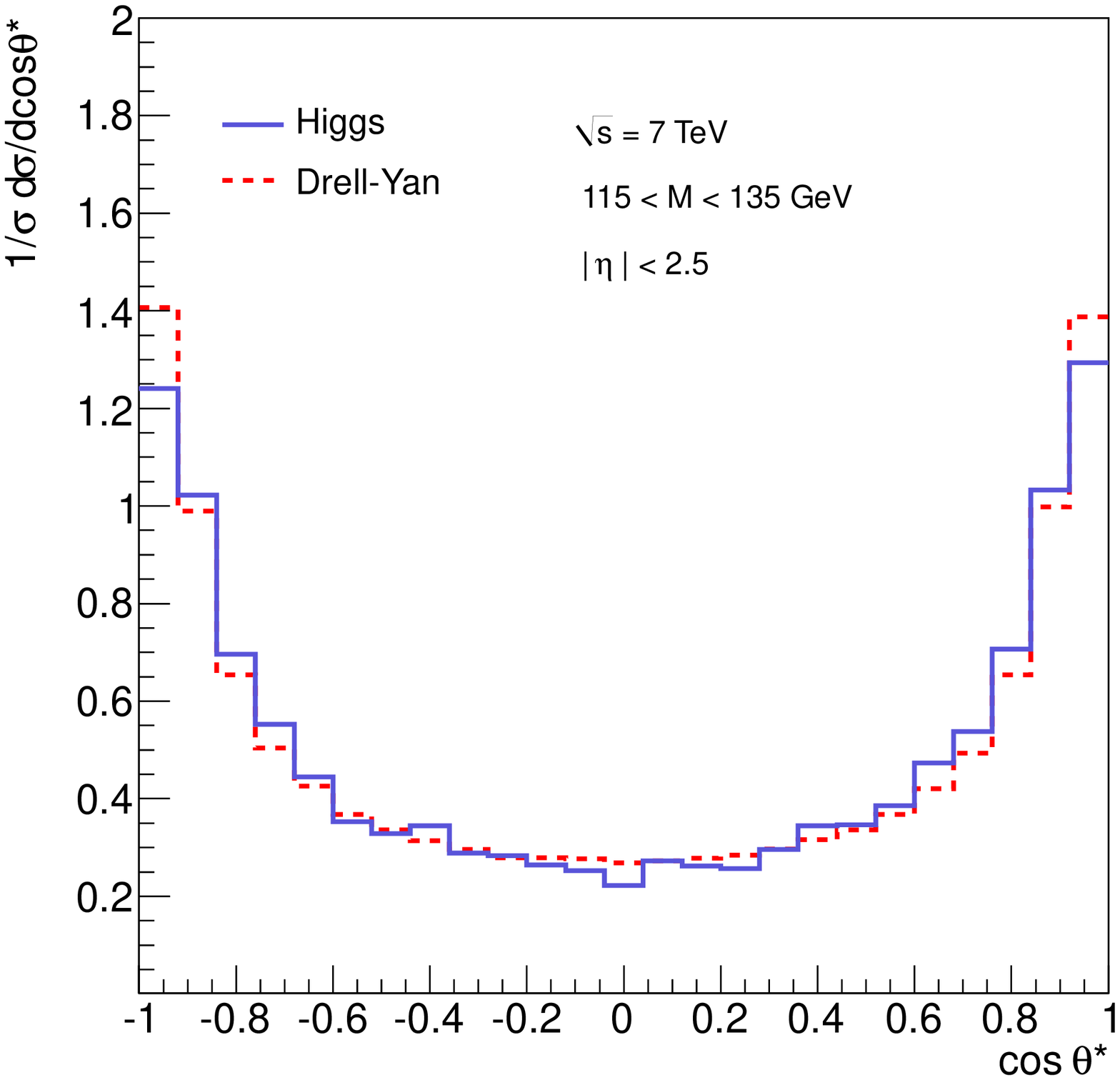}
\includegraphics{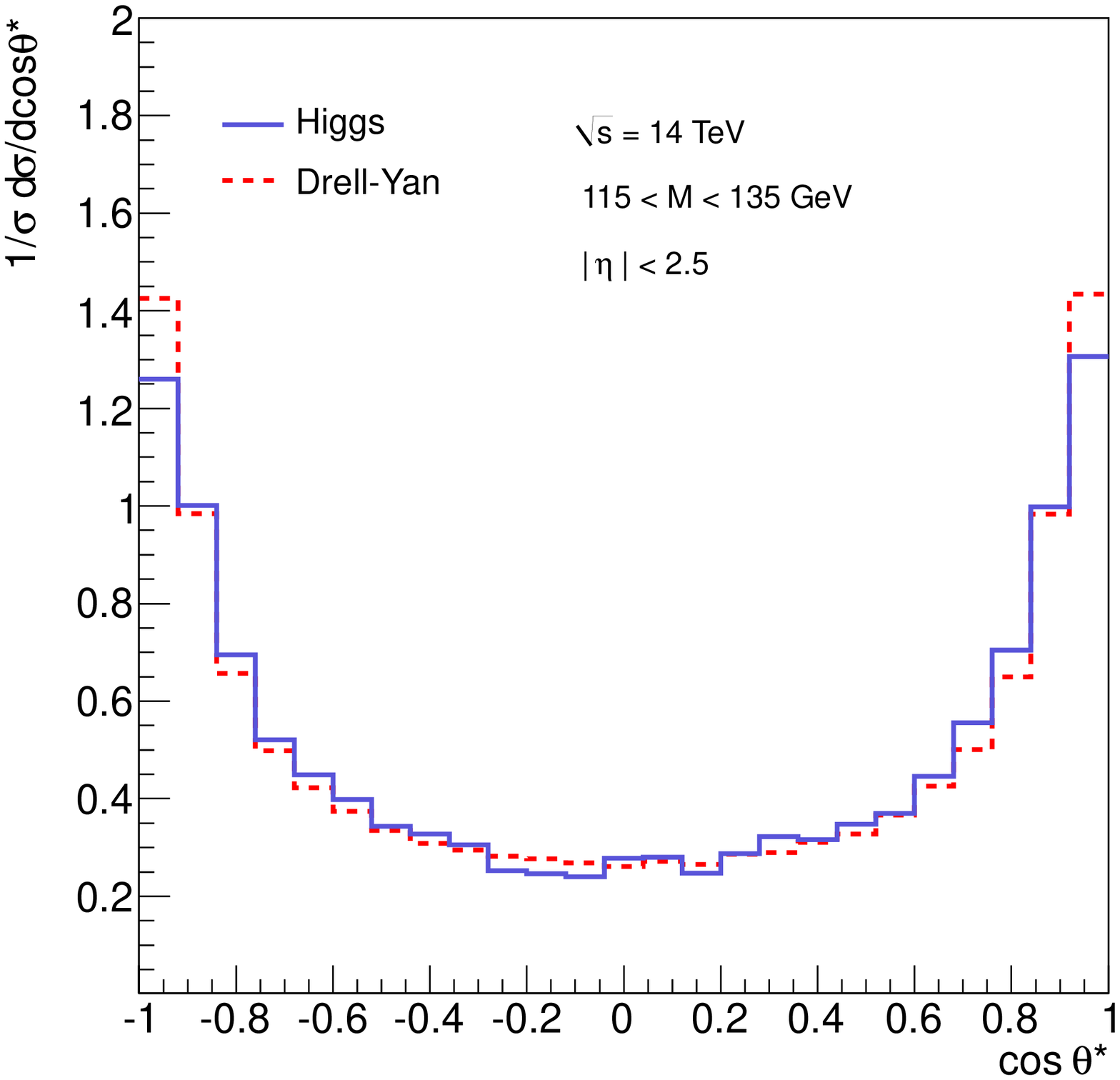}
\caption{\scriptsize  Angular distribution in the center-of-mass scattering angle.} 
\label{fig:fig-costheta}
\end{figure}

In Fig.~\ref{fig:fig-costheta} we  consider   one-jet  production 
 associated  with Higgs and $Z$ bosons, and  show  
 the  differential distributions    in    $\cos \theta^\ast$, 
for 
 jet $p_\perp > 20 $ GeV and boson-jet invariant mass $m$ such that 
200 GeV $ < m < 500 $ GeV. 
The  rise for increasing $\cos \theta^*$ 
reflects the  mechanism described above. 
This large $\cos \theta^*$   power counting   is the 
basic  reason why the 
difference between Higgs and DY in the low-$p_\perp$   region of 
Figs.~\ref{fig:fig-pt}  and~\ref{fig:fig-pt-ratio}  gives a 
measurement of   higher-loop  radiative contributions.
Further effects from higher-order color 
emission may be analyzed  via  angular correlations in the 
boson-jet azimuthal plane   in the laboratory frame.

In summary,  this paper  points  out that  a program of QCD measurements   
 can be carried out in the high-luminosity phase at the LHC, 
  using the  Higgs boson as a  gluon trigger. By measuring 
systematically differences between Higgs and Drell-Yan differential distributions 
for masses around 125 GeV the effects of pile-up largely cancel. 
Such measurements allow one to access experimentally, for the first time, gluon 
 fusion  processes at high mass scales  via a 
color-singlet current.  
Detailed studies are warranted to investigate quantitatively 
the reduction of pile-up contributions in  different channels; 
 the optimal Higgs channels to 
access gluon fusion by suppressing vector boson fusion and 
quark  annihilation;  the required Higgs statistics. 
The    observables  discussed in this paper 
illustrate  that this program spans a broad range of physics issues 
on strong interactions, from soft gluon dynamics showing up 
in the ratio of Higgs to DY low-$p_\perp$ 
spectra, to underlying events and multiple parton interactions 
associated with gluonic showers, 
to hard-QCD  contributions  in    large-$p_\perp$   spectra and 
angular distributions for boson + jet production. 
These  angular distributions  
in particular may be used to test the spin of the Higgs at the level of 
production processes.

{\bf Acknowledgments}.  FH gratefully acknowledges the hospitality and support   of  the 
Terascale Physics Helmholtz Alliance and DESY.


\begin{thebibliography}{999}
\vspace*{-0.25cm}

\bibitem{hgs-obs}    
                ATLAS Coll. (G.~Aad et al.),     Phys.\  Lett.\  B{\bf  716}  (2012)  1; 
         CMS Coll.  (S.~Chatrchyan et al.),   Phys.\  Lett.\  B{\bf  716}  (2012)  30.



\bibitem{eurstra}
           ATLAS and CMS Submission to ``European Strategy for Particle Physics",  Krakow, 
           September 2012. 


\bibitem{eur-qcd}
           CMS Coll.,    ``QCD at the         Extremes",  contribution   to  
           ``European Strategy for Particle Physics",   Krakow,    September 2012. 


\bibitem{atlas-conf-eps}
            ATLAS Coll., ATLAS note ATLAS-CONF-2013-072. 

\bibitem{hin-novaes}
            I.~Hinchliffe and S.F.~Novaes, 
              Phys.\  Rev.\  D{\bf 38} (1988)   3475. 
          

\bibitem{jcc-fh-00}
                J.C.~Collins and F.~Hautmann,      
                Phys.\ Lett.\ B{\bf 472} (2000) 129. 

\bibitem{powhegcit}
               S.~Frixione,   P.~Nason and C.~Oleari,         JHEP {\bf 0711}  (2007)   070. 

\bibitem{pythia6} 
              T.~Sj{\" o}strand, S.~Mrenna and P.~Skands, JHEP 
         {\bf 0605} (2006) 026. 
 

\bibitem{hgs02}
                 F.~Hautmann,          Phys.\ Lett.\ B {\bf 535} (2002) 159. 

\bibitem{mantry} 
           S.~Mantry and F.~Petriello,    
            Phys.\  Rev.\  D{\bf 81} (2010)   093007; 
           Phys.\  Rev.\  D{\bf 83} (2011)   053007.  

 
\bibitem{cms-top}
            CMS Coll.,  CMS-PAS-TOP-12-028; 
            Eur.\  Phys.\   J.\  C {\bf 73} (2013) 2339;   
            ATLAS Coll., ATLAS-CONF-2012-149. 

\bibitem{plaetzer}
             S.~Pl{\" a}tzer, arXiv:1307.0774 [hep-ph]; 
             arXiv:1211.5467 [hep-ph]. 

\bibitem{prestel} 
            L.~L{\" o}nnblad and S.~Prestel,   
               arXiv:1211.4827 [hep-ph];  
       arXiv:1211.7278 [hep-ph]. 

\bibitem{doksjohgs}
           Yu.L.~Dokshitzer, V.A.~Khoze and T.~Sj{\" o}strand, 
             Phys.\ Lett.\ B {\bf 274} (1992) 116. 

\bibitem{skands}
     P.Z.~Skands,        Phys.\  Rev.\  D{\bf 82} (2010)   074018.   


\bibitem{zijl}
          T.~Sj{\" o}strand and M.~van Zijl,   
          Phys.\  Rev.\  D{\bf 36} (1987)   2019. 

\bibitem{tunes} 
           R.D.~Field, arXiv:1010.3558 [hep-ph].   

\bibitem{lonn-sjo} 
          L.~L{\" o}nnblad and M.~Sj{\" o}dahl,
         JHEP {\bf 0402}  (2004)  042; 
            JHEP {\bf 0505}  (2005)  038. 


\bibitem{deak1006} 
         M.~Deak et al.,    arXiv:1006.5401 [hep-ph].   

\bibitem{cms-ue} 
          CMS  Coll.  (S.~Chatrchyan et al.),        JHEP {\bf 1109}  (2011)   109. 

\bibitem{epjc12}
      M.~Deak et al., 
       Eur.\  Phys.\   J.\  C {\bf 72} (2012) 1982.    

\bibitem{eksbook}
      R.K.~Ellis, W.J.~Stirling and B.R.~Webber,  
       {\em QCD and collider physics}, CUP 1996. 

\end{thebibliography}
\end{document}